# Phenomenological Description of a Giant Temperature Hysteresis of the Ultrasound Velocity and the Internal Friction in Lanthanum Manganite

A. P. Saiko* and S. A. Markevich

*Institute of Solid State and Semiconductor Physics, National Academy of Sciences of Belarus, Minsk, 220072 Belarus*
*e-mail: saiko@ifttp.bas-net.by

We propose an explanation for the experimentally observed [1] giant temperature hysteresis of the ultrasound velocity and the internal friction in single crystals of lanthanum manganite ($La_{0.8}Sr_{0.2}MnO_3$). The effect is interpreted within the framework of a phenomenological model based on the notion of two coexisting sublattices of the oxygen octahedra performing cooperative tilting-rotational oscillations in bistable potential fields.

Manganese perovskites $R_{1-x}A_xMnO_3$ ($R$ = La or another rare earth element; $A$ = Ca, Sr, Ba, …) have recently drawn much attention due to the phenomenon of colossal magnetoresistance (CMR) that makes these compounds promising materials for magnetic sensors and data-reading devices. Now it is commonly accepted that an important role in the formation of a CMR response in manganites is played by the interaction of a crystalline lattice with the electron and spin degrees of freedom. However, despite a very large body of experimental data, there are still many open questions related to the quantitative description of the metal–dielectric transition, phase coexistence, and some features in the elastic properties of these compounds.

Recently [1], temperature hysteresis of a giant width (extending over a temperature interval from 80 to 350 K) of the longitudinal ultrasonic wave velocity and the internal friction was observed in a single crystal of lanthanum manganite $La_{0.8}Sr_{0.2}MnO_3$ possessing CMR properties. According to this, the sound velocity in a sample on heating from 80 to 105 K is significantly lower than that on cooling in the same temperature interval, while in the 105–350 K range, a difference in the sound velocities is several times smaller. The curve of the internal friction exhibits a sharp peak at 105 K (on heating) and less pronounced peaks at 350 (on heating) and 80 K (on cooling).

The presence of a giant temperature hysteresis in the elastic properties of lanthanum manganite can be explained within the framework of the phenomenological model of a correlated bistable sublattice [2].

Consider a crystal lattice with a polyatomic basis, in which atoms (ar atomic groups) of one sort perform optical oscillations in an asymmetric double-well potential formed in the core (matrix lattice) field. If the motion of such atoms is strongly correlated and possesses a cooperative character (e.g., due to a long-range order), this will suppress the fluctuational over-barrier transitions between the potential wells. For this reason, and because of the potential asymmetry, the atomic ensemble under consideration may occur in metastable states forming a bistable sublattice. On heating, the sublattice exhibits evolution from oscillations in the global minimum to the overbarrier oscillations with almost excluded slower component (transitions from the global minimum of the seeding potential to the local minimum and vice versa).

It is suggested that octahedra of the perovskite structure of lanthanum manganite, with oxygen ions at the vertices and manganese ions at the center, perform tilting-rotational motions in the double-well asymmetric potential. These oscillations of the octahedra are strongly correlated because of the strong interaction between the charge, spin, and lattice degrees of freedom characteristic of this compound. A transition from the global to metastable minimum of the free energy on heating corresponds to the appearance of rhombohedral distortions in the initially orthorhombic low-temperature structure. There are two types of bistable sublattices of the oxygen octahedra with different positions of metastable minima of the seeding potentials (Fig. 2), which is evidenced by (i) the appearance of experimentally observed rhombohedral component in the orthorhombic phase in a sample heated from zero temperature and (ii) the absence of orthorhombic inclusions on cooling from $T > 350$ K down to 105 K (the two types of bistable sublattices can be related to the presence of two kinds of positive ions, $La^{3+}$ and $Sr^{2+}$). Bistable oscillations of the oxygen octahedra modulate frequencies of the vibrational spectrum of the matrix lattice,

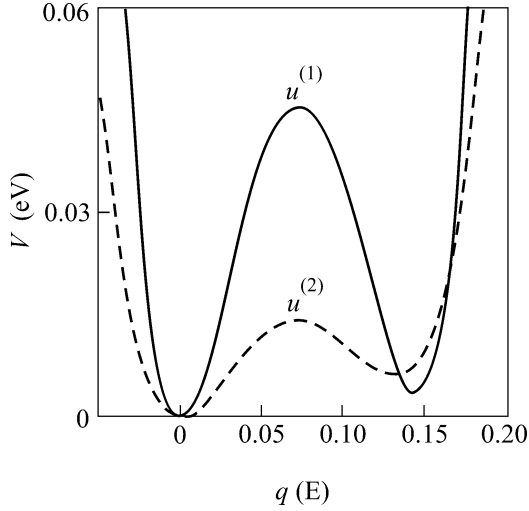

**Fig. 1.** Double-well seeding potentials for the two sublattices of oxygen octahedra: $u^{(1)} = 0.045$ eV (potential barrier height); $q_1^{(1)} = 0.073$ Å and $q_2^{(1)} = 0.144$ Å (coordinates of the maximum and metastable minimum of the seeding potential for the first sublattice); the corresponding values for the second sublattice are $u^{(2)} = 0.014$ eV, $q_1^{(2)} = 0.073$ Å, $q_2^{(2)} = 0.132$ Å.

which is manifested by anomalies in the elastic properties.

The theory of a temperature hysteresis of the elastic properties of lanthanum manganite can be constructed based on the following phenomenological model using a spin Hamiltonian (reduced to a single octahedron) of the type

$$H = H_h + H_{anh} + H_{int}. \quad (1)$$

Here, the first term, $H_h$, represents the matrix lattice Hamiltonian and is selected in the form of a Hamiltonian of the set of harmonic oscillators with the parameters normalized to the empirical values of lattice constants of the compound under consideration. The second term, $H_{anh}$, is an effective Hamiltonian of the oxygen octahedra performing tilting-rotational oscillations in the asymmetric double-well potentials of the two types (Fig. 1):

$$V^{(i)} = \frac{\alpha^{(i)}}{2}(q^{(i)})^2 - \frac{\beta^{(i)}}{3}(q^{(i)})^3 + \frac{\gamma^{(i)}}{4}(q^{(i)})^4, \quad (2)$$

where $q^{(i)}$ are the generalized coordinates describing the vibrational motion of the octahedra in a potential with the higher (lower) barrier and the deep (shallow) metastable minimum, $i = 1$ ($i = 2$); $\alpha^{(i)}$, $\beta^{(i)}$, and $\gamma^{(i)}$ (>0) are the parameters.

Interaction between the lattice oscillators and the nonlinear oscillations of the oxygen octahedra is

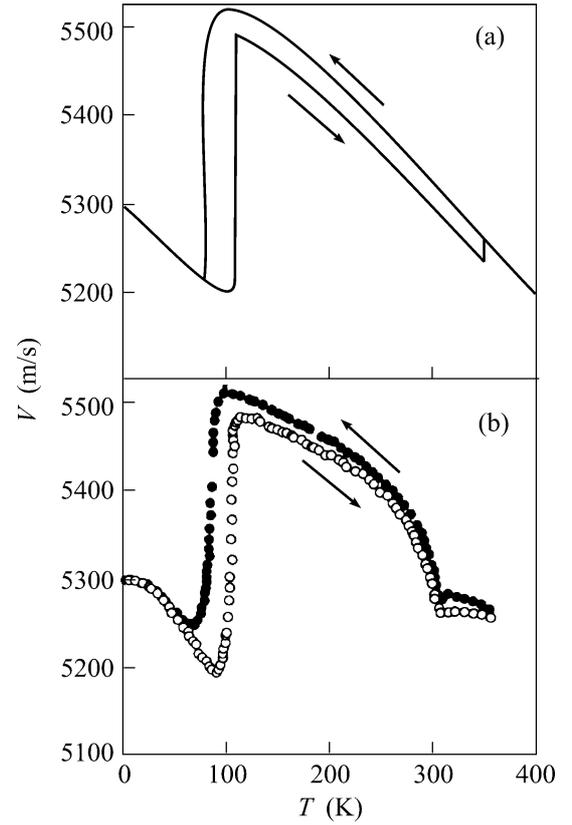

**Fig. 2.** Temperature dependence of the ultrasonic wave velocity: (a) calculated using formula (4) for a crystal with two bistable sublattices coupled by a quartet interaction (5); formula (4) was corrected by adding a contribution from the matrix sublattice, $A-BT$, where $A = 2996$ ms$^{-1}$ and $B = 1.58$ ms$^{-1}$ K$^{-1}$ (determined from experimental data [1]); $\lambda_{kk'}^{(1)}/\mu_k\omega_k^2 = 2$ Å$^{-2}$ and $\lambda_{kk'}^{(2)}/\mu_k\omega_k^2 = 30$ Å$^{-2}$; $u^{(1)} = 0.045$ eV, $q_1^{(1)} = 0.073$ Å, $q_2^{(1)} = 0.144$ Å, $u^{(2)} = 0.014$ eV, $q_1^{(2)} = 0.073$ Å, $q_2^{(2)} = 0.132$ Å; (b) experimental velocity of ultrasound measured for La$_{0.8}$Sr$_{0.2}$MnO$_3$ [1].

described by the third term in Eq. (1), which is selected in the following form:

$$H_{int} = \sum_i (q^{(i)})^2 \sum_{k,k'} (\lambda_{kk'}^{(i)})^2 x_k x_{k'}, \quad (3)$$

where $x_k$ is the displacement of the $k$th oscillator (mode) and $\lambda_{kk'}^{(i)}$ are the force constants of the bonds. An analysis shows that the effective frequency $\varepsilon_k$ of the longwave phonons (or the sound velocity) in this model significantly depends on the temperature and exhibits temperature hysteresis. This is related to a temperature-dependent distortion of the asymmetric potentials (2) and the occupation of metastable states of the oxygen octahedra.

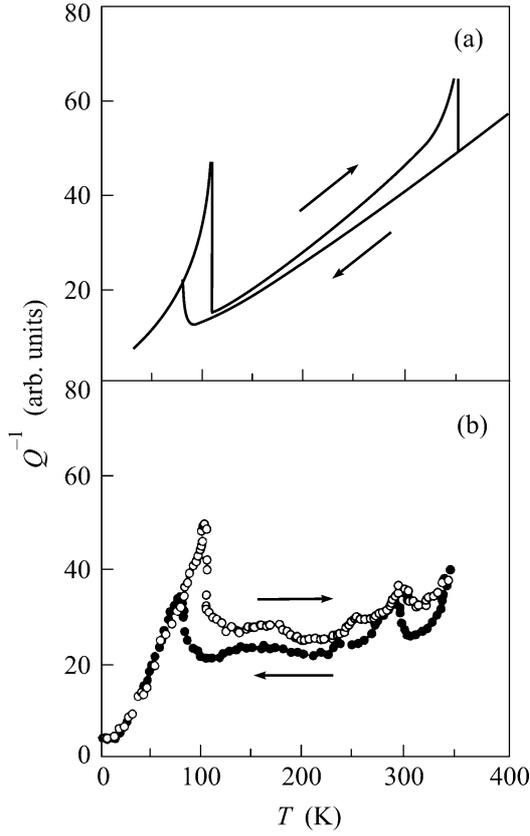

**Fig. 3.** Temperature dependence of the internal friction: (a) calculated for a crystal with two bistable sublattices using formula (7) with $u^{(1)} = 0.045$ eV, $q_1^{(1)} = 0.073$ Å, $q_2^{(1)} = 0.144$ Å, $u^{(2)} = 0.014$ eV, $q_1^{(2)} = 0.073$ Å, $q_2^{(2)} = 0.132$ Å (the contribution of the matrix sublattice is subtracted as background); (b) experimental internal friction curve measured for $La_{0.8}Sr_{0.2}MnO_3$ [1].

The velocity $v$ of an ultrasonic wave is determined from the temperature-dependent frequency $\varepsilon_k$ renormalized by the interaction $H_{int}$:

$$v(T) \sim \varepsilon_k(T) = \omega_k \left[ 1 + \sum_i \frac{\lambda_{kk}^{(i)}}{\mu_k \omega_k^2} (\sigma^{(i)} + \langle q^{(i)} \rangle^2) \right], \quad (4)$$

where $k$ is the wavevector of the mode for which the ultrasonic measurements are performed; $\omega_k$ and $\mu_k$ are the seeding frequency and the mass of the $k$th oscillator in the matrix lattice, respectively; $\langle q^{(i)} \rangle$ and $\sigma^{(i)} = \langle (q^{(i)} - \langle q^{(i)} \rangle)^2 \rangle$ are the statistical mean values of the generalized coordinates and the corresponding dispersions, determined from a system of self-consistent equations,

$$(\beta^{(i)} - 3\gamma^{(i)} \langle q^{(i)} \rangle) \sigma^{(i)}$$
$$= \alpha^{(i)} \langle q^{(i)} \rangle - \beta^{(i)} \langle q^{(i)} \rangle^2 + \gamma^{(i)} \langle q^{(i)} \rangle^3,$$

$$\sigma^{(i)} = \frac{1}{2m\Omega^{(i)}} \coth \frac{\Omega^{(i)}}{2\theta}, \quad (5)$$

where

$$(\Omega^{(i)})^2 = \frac{1}{m} [\alpha^{(i)} - 2\beta^{(i)} \langle q^{(i)} \rangle + 3\gamma^{(i)} (\sigma^{(i)} + \langle q^{(i)} \rangle^2)]$$

are the frequencies of oscillations of the octahedra (the temperature dependences of these frequencies and the $\langle q^{(i)} \rangle$ and $\sigma^{(i)}$ values exhibit hysteresis) and $\theta = k_B T$. The quantum-statistical expressions for $\sigma^{(i)}$ (5) in the regime of strong correlations can be replaced by the corresponding classical limits, $\sigma^{(i)} = \theta/m(\Omega^{(i)})^2$, since the cooperative behavior of the oxygen octahedra extends over several unit cells and, hence, the quantum fluctuations can be ignored in the calculation of means.

Apparently, the temperature hysteresis of the elastic constants of the matrix lattice in the temperature interval from 105 to 350 K will be provided by the cooperative oscillations of octahedra in the anharmonic potential with a higher barrier, while the behavior in the 80–105 K interval will be determined by the cooperative oscillations of octahedra in the potential with a lower barrier. The relative number of octahedra performing oscillations in the anharmonic potential of the first type can be estimated from the experimental data as a fraction of the orthorhombic phase in the rhombohedral phase at $T > 105$ K (this fraction amounts to about 6% [1]). Some other experimental data, such as the relative difference between the elastic characteristics observed on heating and cooling in the temperature interval of bistability and the temperature range of hysteresis, allow obtaining realistic estimates for the parameters of bistable potentials modeling the correlated motions of the oxygen octahedra: $u^{(1)} = 0.045$ eV (potential barrier height); $q_1^{(1)} = 0.073$ Å and $q_2^{(1)} = 0.144$ Å (coordinates of the maximum and metastable minimum of the seeding potential for the first sublattice); the corresponding values for the second sublattice are $u^{(2)} = 0.014$ eV, $q_1^{(2)} = 0.073$ Å, and $q_2^{(2)} = 0.132$ Å; $\lambda_{kk}^{(1)}/\mu_k \omega_k^2 = 2$ Å$^{-2}$ and $\lambda_{kk}^{(2)}/\mu_k \omega_k^2 = 30$ Å$^{-2}$ (normalized constants of coupling of the first and second sublattices to the matrix lattice).

Figure 2a shows a plot of the ultrasonic wave velocity calculated using formula (4). As can be seen, the theoretical temperature dependence well describes the experimental behavior depicted in Fig. 2b. Indeed, in the region of hysteresis, the higher velocities are observed on cooling and the lower, on heating. In the heating mode, there is a sharp increase in $v$ in the region of 105 K, which is related to the appearance of rhombohedral distortions during the tilting-rotational oscillations of correlated octahedra in the bistable potential with a lover barrier $u^{(2)}$. A lower magnitude of the hysteresis in the interval from 105 to 350 K is related to a smaller contribution to the renormalized phonon frequencies from the oxygen octahedra moving in the bistable potential with a higher barrier $u^{(1)}$ (see the legend to Fig. 2). Thus, for reasonable values of the model parameters, both the magnitude and the characteristic temperature intervals (80–105 K and 105–350 K) of the hysteresis, and the direction of traversing this loop in the heating–cooling cycle agree with experiment.

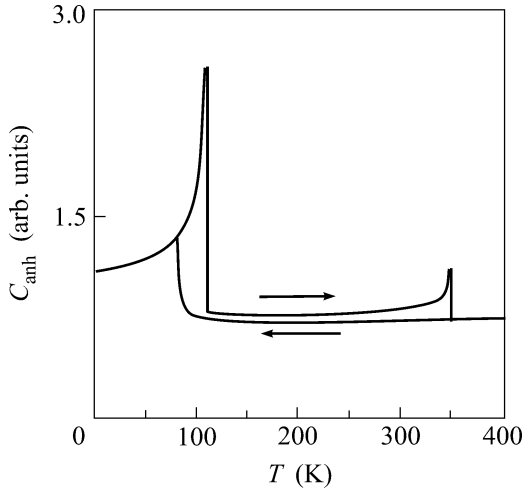

**Fig. 4.** Temperature dependence of the total heat capacity calculated for two bistable sublattices with the parameters (1) $u^{(1)} = 0.045$ eV, $q_1^{(1)} = 0.073$ Å, $q_2^{(1)} = 0.144$ Å and (2) $u^{(2)} = 0.014$ eV, $q_1^{(2)} = 0.073$ Å, $q_2^{(2)} = 0.132$ Å. The weight contribution of the first sublattice is 10% relative total of the second sublattice.

For an ultrasonic wave with a frequency of $\omega_k =$ 80 kHz, we have $\omega_k \tau_{ph} \ll 1$ ($\tau_{ph}$ is the lifetime of thermal phonons). For this reason, a comparison of the theoretical temperature dependence of the viscous friction (ultrasound damping) $Q^{-1}$ with the experimental curve (Fig. 3) was performed using a relation [3]

$$Q^{-1} \sim T C_{anh}. \tag{6}$$

Here, $C_{anh}$ is the heat capacity related to the excitation of anharmonic oscillations in the oxygen octahedra (in Fig. 3, the Debye contribution of the matrix sublattice is subtracted as background), which is calculated using the average energy of these oscillations:

$$\langle H_{anh} \rangle = \sum_{i=1,2} [\alpha^{(i)} \langle q^{(i)} \rangle^2 / 2 - \beta^{(i)} \langle q^{(i)} \rangle^3 / 3 \\ + \gamma^{(i)} \langle q^{(i)} \rangle^4 / 4 + m(\Omega^{(i)})^2 \sigma^{(i)} - 3\gamma^{(i)}(\sigma^{(i)})^2 / 4]. \tag{7}$$

Figure 4 shows the heat capacity $C_{anh}$ calculated (in the classical limit) for the two sublattices of oxygen octahedra. The peaks on the $Q^{-1}(T)$ curve are obviously correlated with the behavior of $C_{anh}(T)$ (cf. Figs. 3 and 4). A more pronounced increase in the internal friction with the temperature in the interval from 105 to 350 K on the calculated dependence as compared to the experimental curve can be related to the temperature-dependent coefficients [3] entering into the formula for $Q^{-1}$, which are difficult to calculate exactly.

Thus, the temperature hysteresis of a giant width in the ultrasound velocity and the internal friction observed in a single crystal of lanthanum manganite $La_{0.8}Sr_{0.2}MnO_3$ can be explained by the cooperative tilting-rotational motions of oxygen octahedra in the bistable potentials of two types (differing by the barrier heights). Interaction of the metastable states of the octahedra with the longwave phonons leads to a temperature-dependent bistable renormalization of the phonon frequencies and, hence, of the ultrasonic wave velocity. The anharmonic contribution from the two oxygen sublattices to the heat capacity, which has a hysteresis character with peaks at the boundaries of the temperature intervals of bistability, accounts for the corresponding temperature behavior of the viscous friction. The calculated and experimental dependences show a good semiquantitative agreement for realistic values of the model parameters.